\documentclass[fleqn,usenatbib]{mnras}

\usepackage{newtxtext,newtxmath}

\usepackage[T1]{fontenc}

\DeclareRobustCommand{\VAN}[3]{#2}
\let\VANthebibliography\thebibliography
\def\thebibliography{\DeclareRobustCommand{\VAN}[3]{##3}\VANthebibliography}

\usepackage{graphicx}	
\usepackage{amsmath}	
\usepackage{textgreek}

\title[Developing Daytime LIF Observation Methods]{Extending Lunar Impact Flash Observations into the Daytime with Short-Wave Infrared}

\author[D.~Sheward et al.]{
D.~Sheward,$^{1}$\thanks{E-mail: djs22@aber.ac.uk (DS)}
M.~Delbo,$^{2}$
C.~Avdellidou,$^{2}$
A.~Cook,$^{1}$
P.~Lognnon\'e,$^{3}$
E.~Munaibari,$^{2}$
L.~Zanatta,$^{4}$
\newauthor
A.~Mercatali,$^{4}$
S.~Delbo,$^{5}$
and P.~Tanga$^{2}$\\
$^{1}$Institute of Maths, Physics and Computer Science, Abersytwyth University, Aberystwyth, United Kingdom, SY23 3FL\\
$^{2}$Universit\'e C\^ote d'Azur, CNRS--Lagrange, Observatoire de la C\^ote d'Azur, CS 34229 -- F 06304 NICE Cedex 4, France\\
$^{3}$Universit\'e Paris Cit\'e, Institut de Physique du Globe de Paris, CNRS, Paris, \^Ile-de-France, FR\\
$^{4}$Setzione di Ricerca Luna dell'Unione Astrofili Italiani, Italy\\
$^{5}$Azienda Servizi Industriali (ASI), Via Konrad Adenauer, 22 - 15067 Novi Ligure (AL), Italy}

\date{Accepted XXX. Received YYY; in original form ZZZ}

\pubyear{2023}

\begin{document}
\label{firstpage}
\pagerange{\pageref{firstpage}--\pageref{lastpage}}
\maketitle

\begin{abstract}
Lunar impact flash (LIF) observations typically occur in R, I, or unfiltered light, and are only possible during night, targeting the night side of a 10-60\% illumination Moon, while >10\textdegree above the observers horizon. This severely limits the potential to observe, and therefore the number of lower occurrence, high energy impacts observed is reduced. By shifting from the typically used wavelengths to the J-Band Short-Wave Infrared, the greater spectral radiance for the most common temperature (2750~K) of LIFs and darker skies at these wavelengths enables LIF monitoring to occur during the daytime, and at greater lunar illumination phases than currently possible. Using a 40.0~cm f/4.5 Newtonian reflector with Ninox 640SU camera and J-band filter, we observed several stars and lunar nightside at various times to assess the theoretical limits of the system. We then performed LIF observations during both day and night to maximise the chances of observing a confirmed LIF to verify the methods. We detected 61 >5\textsigma{ }events, from which 33 candidate LIF events could not be discounted as false positives. One event was confirmed by multi-frame detection, and by independent observers observing in visible light. While this LIF was observed during the night, the observed signal can be used to calculate the equivalent Signal-to-Noise ratio for a similar daytime event. The threshold for daylight LIF detection was found to be between J$_{mag}$=+3.4$\pm$0.18 and J$_{mag}$=+5.6$\pm$0.18 (V$_{mag}$=+4.5 and V$_{mag}$=+6.7 respectively at 2750~K). This represents an increase in opportunity to observe LIFs by almost 500\%.
\end{abstract}



\section{Introduction} \label{sec:intro}

    Due to the shared meteoroid environment of the Earth and Moon, the same distribution of material, from the same sources, impact both the lunar surface and our planet. On Earth, the smaller (sub-meter scale) impactors ablate in the atmosphere as meteors and never reach the ground. In the atmosphereless environment of the Moon, this material instead impacts the lunar surface with its full kinetic energy. 
    
    During the impact, this kinetic energy is partitioned to excavate a crater, provide kinetic energy to ejecta, and heat of the surface and ejecta, while a small fraction (<0.5\%) is released as a flash of light known as a lunar impact flash (LIF). Although typical LIFs only last a fraction of a second, they can be observed from Earth using moderately sized telescopes, allowing both amateur and professional astronomers to monitor them. Since the first confirmed observation in 1999~\citep{ortiz1999, ortiz2000}, over 600 LIFs have been reported in literature and LIF databases~\citep{ortiz2002, ortiz2006, ortiz2015, suggs2014, madiedo2014, madiedo2015b, madiedo2018, larbi2015, bonanos2018, zuluaga2020, avdellidou2021, yanagisawa2021, sheward2022}. 

    Using these observations, analyses have been performed to determine parameters of the impact processes~\citep{suggs2014, madiedo2015, avdellidou2021}, as well as the physical properties of the impactors~\citep{avdellidou2019, avdellidou2021}. Dual-camera observations performed using different wavelength filters (R \& I) on each camera have allowed the temperatures of the flashes to be obtained~\citep{bonanos2018, madiedo2018, avdellidou2019, avdellidou2021}.

    While many studies have been performed, there are still parameters which are not strongly constrained. For example, the luminous efficiency, $\eta$, is the proportion of the impactor's kinetic energy $K.E.$ that is converted into luminous energy, E$_{lum}$, and in literature is taken between 10$^{-4}$ and 10$^{-3}$~\citep{bouley2012,suggs2014}. This order of magnitude uncertainty of this vital parameter leads to the estimation of meteoroids mass with also an order of magnitude uncertainty.

    In order to better constrain the luminous efficiency, efforts are now being made to locate the formed impact craters from observed LIFs. By locating the formed craters, valuable ground truth data is obtained with which crater scaling laws can be applied to derive the amount of energy imparted to crater excavation. We can use this value to empirically determine the percentage of the impactors energy that was not released as light, therefore determining how much was released as light. 
    
    In order to facilitate such a study, open source software was developed in order to locate the formed craters using Lunar Reconnaissance Orbiter Narrow Angle Camera (LRO NAC) images~\citep{sheward2022}. As the LRO NAC has a nominal pixel scale of 0.5~m per pixel, craters of only a few meters rim-to-rim diameter will have large uncertainty in their diameter, and therefore are of less value to be used for further studies. These smaller craters are also more difficult to identify within LRO NAC images, as shadows or processing errors can easily hinder their detection. Consequently, larger craters (>10~m) are preferred for twofold reasons; their larger diameter lowers the percentage error from not knowing where within a pixel the rim of the crater lies, and the larger crater and ejecta blanket allows for the crater itself to be more easily detected and identified within LRO NAC images. 

    The monitoring of impact flashes associated to large impacts is also an important goal for future lunar seismic experiments, as this will provide seismic sources generating high signal to noise seismic signals, with known location and time of the seismic source~\citep{lognonne2009}. This enables structure inversions of the crust~\citep[e.g.][]{chenet2006} or upper mantle~\citep{lognonne2003}. See~\cite{lognonne+johnson2015} for a review on Planetary and Lunar seismology and~\cite{yamada2011} for a demonstration of the importance of impacts monitoring for future seismic network on the Moon. 
    
    The issue with a necessity for larger craters, however, is that the higher energy impacts required to form them are less frequent~\citep{suggs2014,avdellidou2021}. Due to the low frequency of the larger impact events, the importance of maximising the hours in which the Moon is observed for LIFs becomes apparent, in order to capture as many of these events as possible. In order to extend the potential observing hours for LIFs, we present here our justification and methodology for observing using the J-band rather than the typically used R- and I-bands to observed LIFs during local daytime. 
    
    In section~\ref{sec:theory} we present the theoretical basis for the technique. In section~\ref{sec:tests} we discuss the telescope, camera, and experimental setup we used to observe, and the tests we performed in order to verify the theoretical basis. In section~\ref{sec:obs} we present the first results from our lunar observations. We discuss the limits of the system we have developed and discuss the technique and how it impacts LIF observations in section~\ref{sec:discuss}.

\section{Theory}\label{sec:theory}
    \subsection{Current Observations}\label{sub:current}
    Currently LIF observations are performed in R- or I-band, or unfiltered visible light. In order to observe LIFs, two moderately sized (>20~cm) or bigger telescopes  with attached cameras are required~\citep{suggs2014}, or a beamsplitter system wherein two cameras can be attached to a single telescope~\citep{xilouris2018}. The use of two cameras simultaneously observing the same scene is important to discriminate real LIFs from false positives, such as cosmic rays passing through the detector. As the flash only lasts for a fraction of a second, it is essential to use a low exposure rate (<~50~msec) with a high frame rate (>~20~Hz) in order to accurately measure the epoch of the event as well as extract the LIF signal from the background. This background can be quite high due to illumination from earthshine, and stray light from the illuminated hemisphere of the Moon.

    Observing sessions are limited to local (observer) night time (solar altitude~<~-18\textdegree), while the Moon is >10\textdegree{ }above the observers horizon, and has a lunar phase between 10-60\%. Observations have to be performed on the lunar night-side, as far away from the illuminated portion of the lunar surface as possible, in order to avoid the scattered sunlight, to achieve a dark enough background to detect the LIF. These requirements severely limit the number of hours per month in which LIF observations can be performed.
    
    In order to confirm observed candidate flashes as true LIFs, they need to satisfy several conditions. Firstly, the candidate flash must be stationary with respect to the lunar surface, however the brightening or dimming of the flash can be present in subsequent frames as the LIF peaks and wanes. If the candidate flash is present in two or more consecutive frames, the event can be confirmed to be a LIF with high certainty. Movement of the candidate between frames would indicate an interstitial object (between the observer and Moon), such as a satellite or aeroplane.
    
    As cosmic rays can present similarly to a single-frame LIFs, if the flash is only present in a single frame a simultaneous secondary observation is required to confirm the event. The appearance of the candidate flash in only one observation implies that either the candidate could be due to a cosmic ray interaction with one of the cameras, or - given that the two cameras are observing in different wavelengths - that the candidate flash has an intensity below the detection limit in one of the cameras. Neither of these cases can be confirmed or distinguished between without the flash appearing in subsequent frames. 

    In order for a LIF to be detectable, its signal needs to be distinguishable above the background noise in the image. To be detected with certainty, the LIF needs a Signal-to-Noise ratio (SNR) greater than 5~\citep{rose1973}. If the SNR~<~5, the signal cannot be identified with certainty above the noise in the image. The SNR for a LIF can be defined as:

    \begin{equation}\label{eq:snr}
        SNR = \frac{S"}{\sqrt{\frac{S"}{G}+N_{ap}\cdot{\sigma_{b}}^2}} \approx \frac{S"}{\sigma_b\cdot \sqrt{N_{ap}}} 
    \end{equation}
    where S" is the sum of the digital numbers of the background-subtracted point spread function within an area with N$_{ap}$ pixels, $\sigma${$_b$ }is the standard deviation of the background, and G is the gain of the camera. S" is defined as: 

    \begin{equation}\label{eq:counts}
        S" = C" - \tilde{B} \cdot N_{ap}
    \end{equation}
    where C" is the sum of the digital numbers of the point spread function within the area N$_{ap}$, and $\tilde{B}$ is an estimation of the background counts per pixel, generally calculated from the average of an annulus around the source.

    As the SNR is strongly determined by the standard deviation of the background, a lower standard deviation allows for the detection of fainter signals. Similarly, in the case where the standard deviation is greater than the contributions from the dark current and read out noise of the detector, the standard deviation of the background is related to the average count of the background (\textsigma$_b \sim \sqrt{B}$), by reducing the average background count the standard deviation will also decrease.

    \subsection{Observations in J-band}
\subsubsection{LIF brightness}
        In 1975-1976, \citet{eichhorn1975,eichhorn1976} performed impact experiments using Van der Graaf accelerators, and adopted the close approximation that the impact flash radiates energy as a black body, with spectral radiance described by Planck's law:

        \begin{equation}\label{eq:planck}
            B_\lambda(\lambda, T) = \frac{2hc^2}{\lambda^5}\frac{1}{e^{\frac{hc}{\lambda{k_B}T}}-1}
        \end{equation}
        where $B_\lambda(\lambda, T)$ is the spectral radiance at the wavelength, $\lambda$, and temperature, T, $h$ is the Planck constant, c is the speed of light, and k$_B$ is the Boltzmann constant. It has since been assumed in literature that LIFs behave as black bodies~\citep{suggs2014, bonanos2018, avdellidou2019, avdellidou2021}. 
        
        The temperature of the impact flashes can then be estimated by fitting a Planck function to multi-wavelength observations of LIFs. The temperature of the first 112 LIF events observed by NELIOTA has been obtained, with a distribution peaking at around $\approx$2750~K~\citep{avdellidou2021}.
        
        The peak wavelength of the spectral radiance of a LIF can be obtained using Wein's displacement law:

        \begin{equation}\label{eq:weins}
            \lambda_{peak} = \frac{b}{T}
        \end{equation}
        where b is a constant of proportionality, equal to 2898~\textmu{m}~K, and T is the temperature in Kelvin. This gives the typical LIF of 2750~K a peak wavelength of $\lambda$=1.05~\textmu{m}. 

        As previously mentioned, typical LIF observations take place in the R- and I-Bands. The effective midpoint wavelength for these bands in the Johnson-Cousins filter system are $\lambda_{R}$=0.66~\textmu{m}, and $\lambda_{I}$=0.81~\textmu{m}; clearly for LIFs which peak around $\lambda$=1.05~\textmu{m}, observing in these bands is sub-optimal.
        
\begin{figure}
            \centering
            \includegraphics[width=\columnwidth]{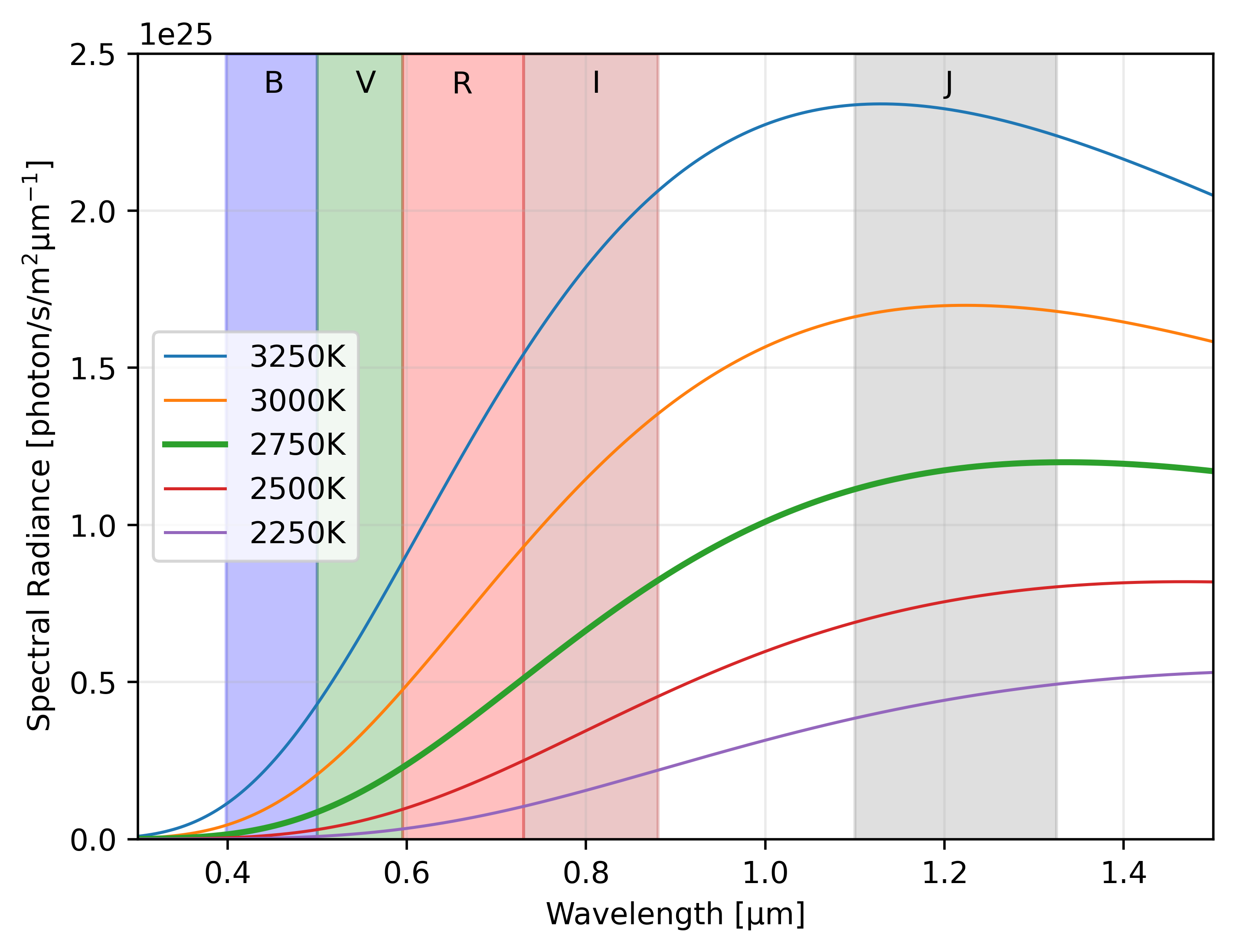}
            \caption{The number of photons released by wavelength for the typical LIF of 2750~K. Temperatures either side of the typical LIF in increments of 250~K are represented by the thinner lines. The coloured bands represent the FWHM bandwidths of the respective filter bandwidths.}
            \label{fig:planck}
        \end{figure}

        As modern LIF observations are typically performed using quantum detectors (see section~\ref{sub:scope}) the number of photons released at each wavelength is responsible for the signal detected by the detectors. Figure~\ref{fig:planck} shows the number of photons released by wavelength for the typical LIF of 2750~K. In J-band approximately 2.6 times more photons are released than in I-band, and 5.3 times more than in R-band, representing an increase in the produced signal, and a much greater possibility to detect LIFs in the J-band.
        
        This translates to a J-band flash of 1.8 magnitudes brighter than in R-band, and 1.04 magnitudes brighter than in I-band, assuming the UBVRI magnitude system based on the number of detected photons.

        While hotter events have the peak wavelength closer to the I-band, at these temperatures the photons emitted by such LIFs are greater in all observed bands, and would therefore still be detectable in the J-band. 

        \subsubsection{Sky Brightness}
        
            Observing in the J-band also provides other advantages over the R- and I-bands for local daytime and twilight observations. The dominant background source of light during local daytime is the atmospheric scattering of solar light, known as dayglow. Dayglow is dominated by the Rayleigh scattering of sunlight through the atmosphere, which is given by the equation: 

            \begin{equation}\label{eq:rayleigh}
                \sigma_R = \frac{8\pi^3}{3}\frac{(m^{2}-1)^2}{\lambda^{4}N^2}
            \end{equation}
            
            \noindent where $\lambda$ is the wavelength of light being scattered, $m$ is the refractive index of the atmospheric gas medium, and $N$ is the number of molecules per unit volume of the atmospheric gas medium, and thus $\sigma_R \propto \lambda^{-4}$.

            This strong dependence on wavelength means that bluer wavelengths have a greater scattering effect, and therefore there is less scattering at longer wavelengths, and consequently a darker sky as observations move into the short-wave infrared. Compared to that of the I-band, in the J-band light is only scattered approximately 29\% as much, which, given the same signal, would equate to an increase in SNR by a factor of 2.4.

            This reduction in scattering also reduces the atmospheric extinction, which allows for observations of the Moon to occur at larger airmass, and therefore lower altitudes, further extending the theoretical observing hours. 
            Consequently by observing in the J-band the theoretical observing period is greatly extended over that of the R- and I-bands. 
            
    \subsubsection{Observing Hours}

        In order to calculate the time that the Moon can be observed for LIFs, we have developed a tool which simulates the sky positions of the sun and Moon, and calculates the illuminated fraction. Running the simulator for observations according to the requirements for V-band, detailed in Section~\ref{sub:current}, over an arbitrary 4 month period we find that from the location of Nice (France) we can observe for $\approx$5.7\% of the time. When adjusting the parameters of the simulation for observing in the J-band for the same 4 month period we can observe for $\approx$27\% of the time. Clearly, observing in the J-band offers a tremendous advantage in terms of time efficiency, giving almost $\times$5 more potential observing hours. This enables the potential for a global network of 6-8 telescopes distributed worldwide at different time zones, to approach continuous lunar observations, allowing future seismic networks to use these impacts for lunar crust tomography~\citep{yamada2011}.

\section{Tests}\label{sec:tests}

In order to prepare for such novel observing mode for LIF we performed some test observations in order to characterise the observing capabilities of a prototype set-up that we developed for this purpose. Observations were obtained from the Mont Gros site of the Observatoire de la C\^ote d'Azur (Nice, France, Latitude=43.7267~deg, Longitude=7.2991~deg).

    \subsection{Equipment}\label{sub:scope}
         The observational setup consisted of a 40.0~cm diameter f/4.4 Newtonian telescope (Skywatcher flex tube 16”), mounted on a custom-made equatorial fork mount, and equipped with a Ninox 640SU InGaAs SWIR camera observing through a J-Band filter. It should be noted that the type of detectors in the this camera are less sensitive than CCDs, however following discussion with ONERA, this should not be an issue. The mount uses a dual stage friction drive on both axes such that no gears are utilized. The axes are driven by high resolution, 1000-steps/turn Oriental Motors steppers, which provide a very smooth and silent operation. Motors are controlled by an Arduino Zero micro-controller, which is connected via an USB-serial link to a host computer (Dell optiplex running Linux Ubuntu). Home build absolute encoders are installed on each axis, which are used for safety limits and pointing the telescope. The concept of these angular encoders and their calibration is described in App.~A. 

         A small, 6~cm in diameter, 41.5~cm focal length refracting telescope, mounted in parallel to the main telescope, is used for auto-guiding on the lit-hemisphere of the Moon. This instrument is equipped with an ASI-ZWO 174 uncooled camera. We developed an \textit{ad hoc} guiding algorithm: each frame (or subframe)  is cross-correlated with a reference frame, the latter taken at the beginiing of each guiding session. The position of the maximum of the cross-correlation function, at subpixel accuracy, is used to calculate an image displacement. The latter is converted in arcseconds and sent to the telescope control system in order to slightly change the position of the motiors to counterbalance the image shift. This method allows us to guide with 0.2-0.4~arcsec RMS even in full daylight.
         
         The Ninox 640SU captures 640$\times$512 16-bit video, and has a pixel pitch of 15~\textmu{m}$\times$\textmu{m}, giving a field-of-view of 18.6'$\times$14.4', covering approximately 35\% of the lunar surface through the described telescope. The camera can operate at up to 90~fps, with a read-out time of 10.2~msec, a dark current noise of <300~e- pix$^{-1}$sec$^{-1}$, and a read-out noise of <98~e- RMS pix$^{-1}$. 

        \begin{figure}
            \centering
            \includegraphics{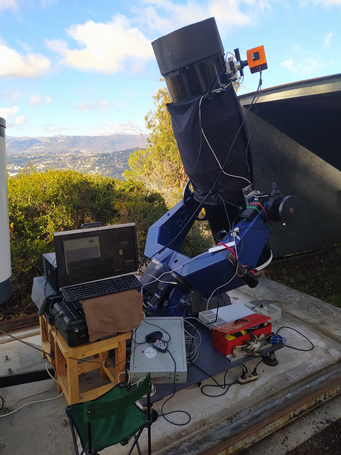}
            \caption{The telescope and camera set-up at Observatoire de la C\^ote d'Azur. Cardboard baffling has been used to occlude stray light from entering through the midsection of the telescope, and likewise on the end of the telescope to ensure the secondary mirror was permanently shaded.}
            \label{fig:scope}
        \end{figure}

    \subsection{Instrument Photometric Calibrations}\label{sub:stars}

        In order to quantify the observational capabilities of the system, several stars were observed for calibration measurements on different days under different and variable weather conditions, and at different airmasses. M- and K-type stars near the path of the Moon were targeted for observation for two reasons; firstly the temperatures of these stars range from 2000-3000~K for M-types, and 3000-5000~K for K-types, giving them similar black body spectra to those of the LIFs~\citep{avdellidou2021}. Secondly, the proximity to the path of the Moon ensures the observations are taken at a similar airmass and light pollution level (the telescope is looking at the sky above the bright city light of Nice downtown) to the LIF observations. Several stars of other spectral types were also observed as targets of opportunity, appearing within the frame of lunar limb observations. The stars observed are summarised in Table~\ref{tab:stars}.

        In order to utilise these star measurements to assess the SNR limits of the system, the star magnitudes needed to first be corrected for the atmospheric extinction. Corrections can be performed using the equation:

        \begin{equation}\label{eq:extinct}
            M_0(\lambda) = -2.5 \log_{10} \left( g\frac{S"}{t}\right) + \kappa(\lambda)\chi + Z_P
        \end{equation}
        
        \noindent where M$_0$ is the exoatmospheric apparent magnitude of the star, S" is the signal as described in Eq.~\ref{eq:counts}, g is the gain of the camera, 0.75~electrons count$^{-1}$, and t is the exposure time. The atmospheric extinction coefficient for the wavelength, $\lambda$, is $\kappa$, $X$ is the airmass, and Z$_P$ is the zero point magnitude - the magnitude which would produce a signal of 1 count per second. 

        We found that applying Eq.~\ref{eq:extinct} directly on the star observations in Tab.~\ref{tab:stars} results in an unreliable solution for \textkappa. This is likely because of the different sky conditions under which these observations were carried out.
        
        In order to calculate $\kappa$ for the J-band, observations were performed of the star $\beta$~Cet every 10 minutes over the course of an evening under good sky conditions, and recording the airmass for each observation. For each observation the star was acquired, and then the telescope slewed to move the star's placement within the image frame. The frames containing the star at location A were then averaged, as were the stars at location B, forming images A and B respectively. A-B subtraction was then performed, which removes the dark current and the majority of the background, leaving a near-zero residual background, and the clean star signals of +A and -B. The absolute of the signals can be averaged to obtain the average signal produced by the star. The signal the star produces for each observation can be converted into an instrument magnitude, M$_i$, with the equation:

        \begin{equation}
            M_i = -2.5\log_{10}(S").
        \end{equation}

        \noindent By plotting the airmass against \textDelta{M}, where:

        \begin{equation}
            \Delta M = M_0 - M_i
        \end{equation}
        
       \noindent the value for $\kappa$ can then be obtained in the form of the slope of a linear fit through the data. By doing so with the data obtained, discarding outliers, we obtained a value of $\kappa$~=~0.112$\pm$0.046 for the J-band, in addition the value of the photometric zero point can also be derived: we found to be Z$_P$~=~20.0$\pm$0.1.
       
       We can now return to the measurement of Tab.~\ref{tab:stars} and we can use Eq.~\ref{eq:extinct} to correct each measurement for extinction and estimate an average Z$_P$-value. We find Z$_P$~=~19.8 with a standard deviation of 0.2~mag. We can thus deduce that the Z$_P$ under good conditions is around Z$_P$=20.0$\pm$0.1 and under average conditions is around Z$_P$=19.8$\pm$0.2. 

        \begin{table*}\label{tab:stars}
            \centering
            \caption{The stars observed for benchmarking the theoretical limits of LIF detection with this experimental setup. The spectral type of the star, and V- and J-band apparent magnitudes are given, obtained from SIMBAD$^{(a)}$. The signal in e$^-$s$^{-1}$ was obtained using the nominal median value for the low gain mode of the camera, 0.75~electrons count$^{-1}$. A is the airmass that the star was observed at.}

       \begin{tabular}{c|c|c|c|c|c|c|c|c|r}
            \hline
            \hline
            Star & Effective & Date & UT Time & A & V-band & J-band & Exp    & Signal \\
            Name & Temp (K)  &      &         &   & Mag    & Mag    & (msec) & (e$^-$s$^{-1}$) \\
            \hline
            Arcturus & 4375 & 2022-12-17 & 10:57:39 & 1.39 & -0.05 & -2.25 & 0.5 & 597813655.50 \\  
            \textalpha Ari  & 4480 & 2023-02-02 & 18:36:24 & 1.15 & 2.01 & 0.06 & 2.5 & 77756220.50 \\  
            \textbeta Cet  & 4797 & 2023-02-02 & 17:30:23 & 2.63 & 2.01 & 0.39 & 2.5 & 58404379.22 \\  
            HD224935 & 3647 & 2022-12-16 & 18:04:39 & 1.55 & 4.41 & 0.67 & 10.0 & 25981983.23 \\  
            HD224062 & 3500 & 2022-12-16 & 18:22:44 & 1.40 & 5.72 & 1.26 & 5.0 & 19680206.85 \\  
            \textTheta Cet & 4660 & 2023-02-02 & 18:23:16 & 2.00 & 3.59 & 1.84 & 5.0 & 12542898.06 \\  
            \textbeta Per & 13000 & 2023-02-02 & 18:42:03 & 1.01 & 2.12 & 2.16 & 20.0 & 8760372.62 \\  
            \textalpha Peg  & 9765 & 2022-12-16 & 17:27:16 & 1.15 & 2.48 & 2.51 & 29.8 & 7000501.61 \\  
            HD224677 &  3773 & 2022-12-16 & 17:44:08 & 1.39 & 6.91 & 3.52 & 29.8 & 3109231.18 \\  
            \textalpha Psc  & 10000$^{(b)}$ & 2023-02-02 & 18:30:48 & 1.47 & 3.75 & 3.75  & 29.8 & 2465487.56 \\  
            HD224331  & 3895 & 2022-12-16 & 18:30:05 & 1.34 & 7.22 & 5.01 & 29.8 & 876108.56 \\  
            HD224346  & 4650 & 2022-12-16 & 18:30:05 & 1.34 & 7.57 & 5.62 & 29.8 & 353020.93 \\  
            HD224382  & 6750 & 2022-12-16 & 18:30:05 & 1.34 & 7.61 & 6.87 & 29.8 & 93601.90 \\  
            \hline 
            
        \end{tabular}
        \newline\noindent\footnotesize{$^{(a)}$ https:$//$simbad.unistra.fr$/$simbad$/$; $^{(b)}$ This system is a binary containing two stars of this temperature.}
        \end{table*}

        Having assessed the instrument photometric zero points, we can estimate instrument’s limiting magnitude for LIF detection. Following well established methods \citep{ortiz1999}, we adopt the concept that a LIF should be detected  5$\sigma$ above the background noise, with a least 6 pixels in order to minimise false positives from the noise characteristic described in \ref{sub:noise}. We considered four background cases that we measured during the 2023-03-27 run, when we observed from about 12:00 until 20:30~UT. During daytime observing the level of the background was measured — pointing the telescope at the non-sunlit hemisphere of the Moon — to be around 24,000 counts, which corresponds to 18,000 electrons in 1~msec of integration time. Since the standard deviation of the background is approximately proportional to the square root of its value, this implies a standard deviation of about 135 electrons. We require a signal to be in at least 6 pixels, 5 times above the background standard deviation, which implies that $135\times\sqrt{6}\times5=1653$~electrons in a msec, or an equivalent signal of $1.65\times10^6$~electrons sec$^{-1}$. Using Eq.~\ref{eq:extinct} with the Z$_P$ and \textkappa{ }values determined before, at airmass 2 we calculate that this is equivalent to J=+4.82$\pm$0.18 mag.

        Immediately after sunset, the sky background varies rapidly from the daylight level to the night time level, the contribution from Earthshine and starlight from the nearby bright hemisphere of the Moon is strongly reduced in the J band compared to visible light bands. We measured background of 8,000 and 13,000 counts with exposure times of 10 and 23~msec respectively during twilight, as well as a 4,000 counts with exposure time of 30~msec when the non-illuminated part of the Moon was observed. Applying the same method for these three further background cases, we obtain limiting magnitudes of +7.82$\pm$0.18, +8.51$\pm$0.18 during twilight and +9.27$\pm$0.18 during the night in good conditions, and decrease by $\approx$0.2 magnitudes during average conditions.

        At exposure times of 1~msec and 10~msec, when including the 10.2~msec camera readout time the frame period is significantly shorter than that of the 30~msec exposure. In order to improve the limiting magnitude for these cases we can co-add frames together to improve the SNR, and therefore the limiting magnitude. As the duration of the most common impact flashes are typically between 33~msec and 66~msec, we can assume to be able to co-add 4 frames of exposure time 1~msec, resulting in a total frame period of 44.8~msec, or 2 frames of 10~msec, resulting in a total frame period of 40.4~msec. As a result of Eq.~\ref{eq:snr}, adding N frames would roughly multiply both the signal and the noise by N, leading to an increase in the SNR by a factor of $\sqrt{N}$. Consequently, the signal needed to achieve an SNR of 5 is decreased by this amount, leading to the detectable limits becoming J=+5.6 for 4 frames at 1~msec, and J=+8.19 for 2 frames at 10~msec. 
        It is to note that during daylight 24,000 counts in a pixel of 1.7x1.7~arcsec corresponds to a sky J=+3~arcsec$^{-2}$. Our telescope is installed at about 372~m above sea level; J-band sky brightness between 1 and 2 magnitudes fainter can be achieved from higher-altitude astronomical sites \citep{hart2014}.

    \subsection{Noise Characterisation}\label{sub:noise}
    
\begin{figure}
            \centering
            \includegraphics[width=\columnwidth]{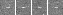}
            \caption{Four examples of the camera noise characteristic that were detected by the LIF identification software.}
            \label{fig:noise}
        \end{figure}

        When observing for LIFs, one potential source of false positives is camera noise. With the Ninox 640SU used in this work there is an intermittent, sporadic noise characteristic without a known cause or pattern. In order to easily disregard the resultant false positives, experiments were performed to characterise the noise of the camera. By running the camera in the same configuration as when observing for LIFs, with the lens cap left on preventing light from entering the camera we can obtain dark frames with similar interference, and similar chance of cosmic ray interactions as during telescope observations on the sky. By running this experiment for 1~hour we obtained 61 events above the 5$\sigma$ detection threshold, three of which appear to be cosmic ray interactions which occurred non-normal to the camera CCD, leaving a trail as it passed through the sensor. The remainder were artifacts of the cameras noise profile, all occurring in a distinctive horizontal pattern, as seen in Fig.~\ref{fig:noise}, which are easily identifiable and therefore easy to discount if registered as a LIF candidate.

\section{LIF Monitoring and Performance}\label{sec:obs}

    The observations used in this work took place over 14 sessions, resulting in just under 40~hours of LIF monitoring. The LIF detection software described in \citet{avdellidou2021} was utilised to process the observations, and identify any potential LIFs. 
    Over this course, 61 LIF candidates were detected 5$\sigma$ above the noise level that could not immediately be discounted as a noise characteristic (see Tab.~\ref{tab:obs}). Once each of these LIF candidates had been examined, and any cosmic rays and interstitial objects moving between subsequent frames are disregarded, this left 33 candidate events as potential LIFs. To further determine which, if any, of these events are true LIFs, the appearance of these events in previous or subsequent frames is examined, and was found that in candidate 45 the flash appears in eight consecutive frames without movement, confirming this event as a true LIF. In addition, its brightness decreases in successive frames, which is expected from a LIF. All other events were single-frame events, which without simultaneous observations are unable to be verified as LIFs. The NELIOTA database was also checked for events coinciding with our observations, with only a single event being detected by NELIOTA during our observations. This event occurred on 2023-03-26 at UT20:25:28.547, and was not detected by us, as the flash occurred outside of field-of-view of our system. Potential reasons for a lack of other NELIOTA events co-occuring with our observations could be due to either NELIOTA not observing, or the small number of hours in which night-time observations can occur by NELIOTA in Greece, while simultaneous daytime observations are taken from Nice, France.   

    \begin{table*}
            \centering
            \caption{Summary of the detections.}
            \label{tab:obs}
            \begin{tabular}{c|c|c|c|c|c|c|c}
            \hline
            \hline
             & &  & Solar & Exp & \# of &  &  \\
           ID & Date & UT Time & Elevation (\textdegree)& (msec) & Frames & J-Mag & Comments \\
            \hline
            1 & 2022-11-26 & 16:28:38.18 & -5.93 & 29.8 & 1 & 7.38&\\
            2 & " & 16:48:40.85 & -9.30 & 29.8 & 1 & 7.89&\\
            3 & " & 16:51:18.83 & -9.74 & 29.8 & 1 & 7.89&\\
            4 & " & 16:54:48.33 & -10.34 & 29.8 & 1 & - & Cosmic Ray\\
            5 & " & 16:55:14.03 & -10.41 & 29.8 & 1 & 8.44&\\
            6 & " & 16:57:21.30 & -10.77 & 29.8 & 1 & 7.83 & Out of Focus, Likely Interstitial Object\\
            7 & " & 16:57:37.68 & -10.82 & 29.8 & 1 & 7.81&\\
            8 & " & 16:57:43.53 & -10.84 & 29.8 & 1 & 7.18 & Out of Focus, Likely Interstitial Object\\
            9 & " & 17:06:30.37 & -12.34 & 29.8 & 1 & 7.76&\\
            10 & " & 17:08:32.42 & -12.70 & 29.8 & 1 & 6.71&\\
            11 & " & 17:08:36.19 & -12.71 & 29.8 & 1 & 7.34&\\
            12 & " & 17:11:04.42 & -13.13 & 29.8 & 1 & 7.58 & Out of Focus, Likely Interstitial Object\\
            13 & " & 17:11:54.22 & -13.28 & 29.8 & 1 & 8.14 & Out of Focus, Likely Interstitial Object\\
            14 & " & 17:12:02.97 & -13.30 & 29.8 & 1 & 7.60&\\
            15 & " & 17:13:26.52 & -13.54 & 29.8 & 1 & 7.48&\\
            16 & " & 17:14:06.14 & -13.66 & 29.8 & 1 & - & Cosmic Ray\\
            17 & 2022-12-01 & 22:47:11.58 & -67.15 & 29.8 & 1 & - & Cosmic Ray\\
            18 & " & 22:51:25.38 & -67.40 & 29.8 & 1 & 7.10&\\
            19 & " & 23:18:23.99 & -68.19 & 29.8 & 1 & - & Cosmic Ray\\
            20 & " & 23:19:53.07 & -68.19 & 29.8 & 1 & 5.93&\\
            21 & 2022-12-18 & 07:15:08.95 & 1.52 & 10.0 & 1 & - & Cosmic Ray\\
            22 & " & 07:46:41.63 & 6.01 & 10.0 & 1 & - & Cosmic Ray\\
            23 & " & 07:04:32.19 & 8.39 & 10.0 & 1 & 4.91&\\
            24 & " & 07:05:28.42 & 8.51 & 10.0 & 1 & 4.81&\\
            25 & 2022-12-27 & 16:22:49.03 & -4.38 & 10.0 & 1 & - & Cosmic Ray\\
            26 & 2022-12-28 & 18:42:23.12 & -28.19 & 29.8 & 1 & - & Cosmic Ray\\
            27 & " & 19:46:52.66 & -39.82 & 29.8 & 1 & 6.57&\\
            28 & 2023-01-25 & 19:31:36.79 & -32.38 & 29.8 & 1 & 7.01&\\
            29 & " & 19:37:02.36 & -33.35 & 29.8 & 1 & 7.28&\\
            30 & 2023-01-26 & 16:42:30.96 & -2.39 & 23.1 & 1 & 5.77&\\
            31 & " & 17:06:36.66 & -6.43 & 29.8 & 1 & - & Cosmic Ray\\
            32 & " & 17:17:42.23 & -8.33 & 29.8 & 1 & - & Cosmic Ray\\
            33 & " & 17:19:38.11 & -8.66 & 29.8 & 1 & - & Interstitial Satellite\\
            34 & " & 17:19:39.79 & -8.66 & 29.8 & 1 & - & Interstitial Satellite\\
            35 & " & 17:22:31.95 & -9.16 & 29.8 & 1 & 7.19&\\
            36 & " & 17:28:00.52 & -10.10 & 29.8 & 1 & - & Cosmic Ray\\
            37 & " & 17:30:48.56 & -10.59 & 29.8 & 1 & 6.80&\\
            38 & " & 17:39:47.22 & -12.16 & 29.8 & 1 & - & Cosmic Ray\\
            39 & " & 17:50:10.25 & -13.98 & 29.8 & 1 & 6.50&\\
            40 & " & 17:53:00.41 & -14.48 & 29.8 & 1 & 7.76&\\
            41 & " & 19:05:22.71 & -27.44 & 29.8 & 1 & 6.61&\\
            42 & " & 19:35:26.15 & -32.86 & 29.8 & 1 & - & Cosmic Ray\\
            43 & " & 20:00:33.15 & -37.36 & 29.8 & 1 & 6.82&\\
            44 & " & 20:16:44.63 & -40.22 & 29.8 & 1 & 6.84&\\
            45 & " & 20:33:30.96 & -43.14 & 29.8 & 17 & 3.19$^{(peak)}$ & Confirmed Impact Flash\\
            46 & " & 20:34:06.72 & -43.24 & 29.8 & 1 & 5.90&\\
            47a & " & 20:44:40.56 & -45.05 & 29.8 & 1 & 6.22 & Two Candidates in Frame 47\\
            47b & " & 20:44:40.56 & -45.05 & 29.8 & 1 & 8.00 & Two Candidates in Frame 47\\
            48 & " & 20:51:48.48 & -46.26 & 29.8 & 1 & 7.34&\\
            49 & 2023-01-30 & 17:03:25.36 & -5.02 & 23.1 & 1 & 6.51&\\
            50 & " & 17:48:43.73 & -12.88 & 23.1 & 1 & - & Cosmic Ray\\
            51 & 2023-03-26 & 18:15:14.27 & -5.38 &20.0&1& 5.85 & \\
            52 & " & 18:15:47.26 & -5.48 &20.0&1&-& Cosmic Ray\\
            53 & " & 18:18:49.43 & -6.02 &20.0&1&-& Cosmic Ray\\
            54 & " & 18:31:39.12 & -8.30 &29.8&1&-& Cosmic Ray\\
            55 & " & 19:05:51.51 & -14.28 &29.8&1&-& Cosmic Ray\\
            56 & " & 19:21:58.34 & -17.03 &29.8&1&7.26&\\
            57 & " & 19:52:12.29 & -22.02 &29.8&1&-& Cosmic Ray\\
            58 & " & 21:10:18.31 & -33.49 &29.8&1&-& Cosmic Ray\\
            59 & " & 20:41:00.56 & -35.56 &29.8&1& 7.22 &\\
            60 & 2023-03-27 & 19:41:57.03 & -20.11 &29.8&1&-& Cosmic Ray\\
            \hline
            \end{tabular}
            \end{table*}

    \subsection{First LIF detection in J-band}
    
\begin{figure}
            \centering
            \includegraphics[width=\columnwidth]{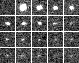}
            \caption{The background-subtracted cropped region of interest for the confirmed flash, event ID 45. The first frame is immediately before the flash, and the subsequent 17 frames that are able to be confirmed above the 5$\sigma$ threshold. The final two frames appear to contain the flash, however as the signal is below the 5$\sigma$ threshold, it cannot be distinguished from noise with certainty.}
            \label{fig:flash}
        \end{figure}

\begin{figure}
            \centering
            \includegraphics[width=\columnwidth]{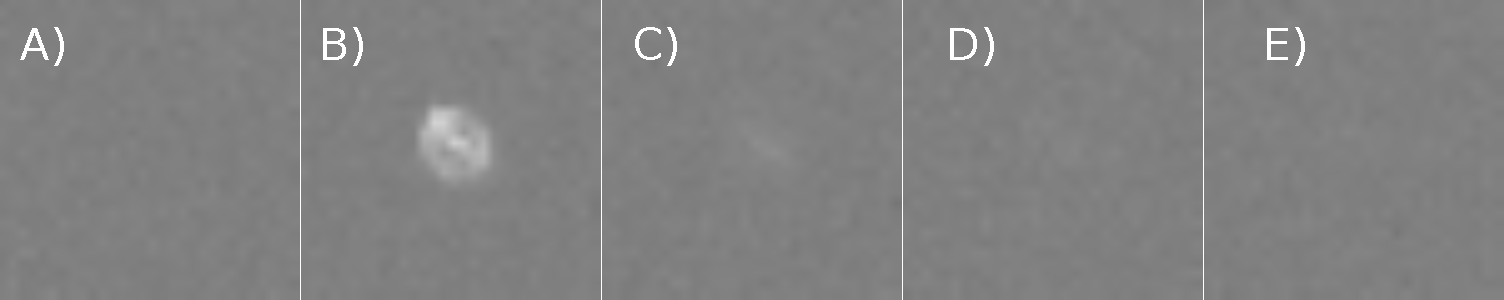}
            \caption{The background-subtracted, cropped region of interest for the confirmed flash, event ID 45, as observed by SdR UAI. Due to the equipment being slightly out of focus, the flash has a shape resembling the telescope mirror. Frames A) and E) contain only the residual background, and frames B), C), and D) contain the flash signal. In frame D) the flash is difficult to discern, as the lack of focus spreads the signal over a wider area than a point source, however is still detectable as SNR~>~5.}
            \label{fig:flash2}
        \end{figure}

        At 20:33:30.96 UT on 2023-01-26, LIF candidate 45 was observed by both the system described in Section~\ref{sub:scope}, and independently by Sezione di Ricerca Luna dell'Astrofili Italiani (SdR UAI) through a 20.0~cm aperture, 100~cm focal length Newtonian telescope, from a site in Melazzo, AL, Italy. The observation from Italy was performed in unfiltered visible light, into an ASI120MM CMOS camera running at 25~fps. The captured visible light flash consisted of three frames above the 5$\sigma$ detection threshold. Fig.~\ref{fig:flash} shows the flash as observed in the J-band, while Fig.~\ref{fig:flash2} shows the flash as seen in the visible.
        
        \subsubsection{Photometry}

        As no stars were observed during the 2023-01-26 observing session, the J-band magnitude of the flash for each frame was calculated using some of the observed calibration stars from Tab.~\ref{tab:stars}. By comparing to the 3 calibration stars closest in airmass to that of the LIF (\textbeta~Cet, \textalpha~Peg, and \texttheta~Cet), and an average can then be taken of the results. The stars were first corrected to the airmass of the LIF, and then aperture photometry was performed in AstroImageJ, using the equation: 
        
        \begin{equation}
            M_{flash} = \frac{1}{3} \sum_{i=1}^{3} M_i - 2.5 \log_{10}\left( \frac{F_{flash}}{F_{i}} \right)
        \end{equation}
        
        \noindent where M$_{flash}$ is the J-band magnitude of the flash, M$_{i}$ are the airmass-corrected J-band magnitudes of the reference stars, and F$_{flash}$ and F$_{i}$ are the signal counts of the flash and average counts of the reference stars respectively. 
        
        \begin{figure}
            \centering
            \includegraphics[width=\columnwidth]{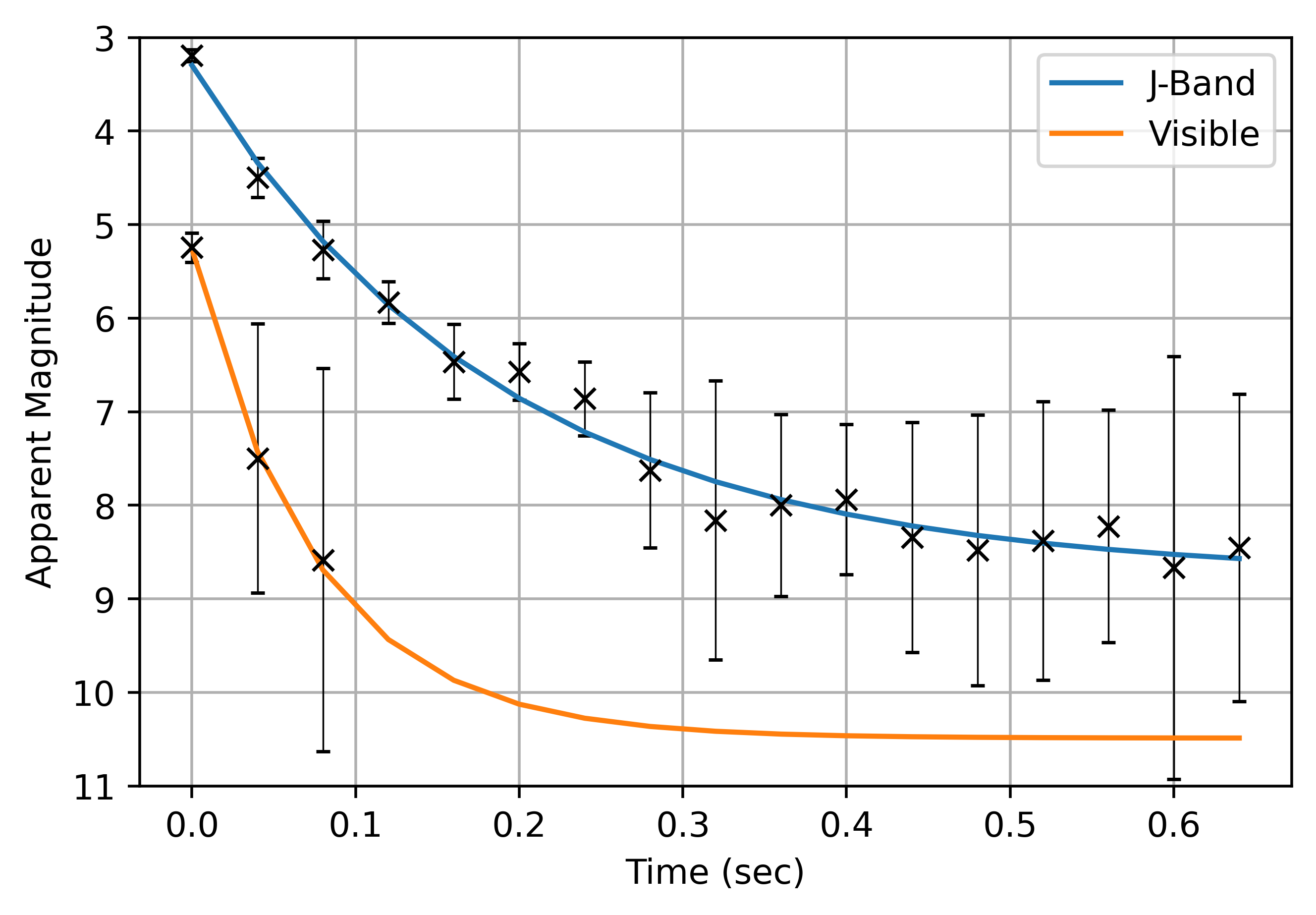}
            \caption{The evolution of event 45 in J-band and unfiltered visible light.}
            \label{fig:mags}
        \end{figure}

        As there were also no stars observed in the SdR UAI observations, aperture photometry was performed using observations of a star taken on a previous night using the same equipment. This gave the flash a J-band peak of Mag$_J$~=~+3.19$\pm$0.18, and visible light peak of Mag$_{Vis}$~=~+5.24$\pm$0.34. The ASI ZWO 120MMs quantum efficiency peaks around 0.55~\textmu{m}. From Fig.~\ref{fig:planck} it can be seen that the ratio of LIF photon spectral density at 1.2 and 0.55~\textmu{m} is approximately a factor of 6, i.e. a difference of about 2 magnitudes between the magnitudes obtained by the two stations. This is very close to our photometric results.
        
        The observed magnitudes and the flash's evolution in both J-band and visible can be seen in Fig.~\ref{fig:mags}. It should be noted that the timestamps in both cameras are not millisecond accurate to GPS time, and therefore sub-frame alignment is not possible. In order to derive the approximate black body temperature of the flash, the assumption is made that the frames are co-occurring.

        From the magnitudes obtained we can calculate the luminous energy, E$_{lum}$ released in the J-band during the impact by first calculating the flux density of each frame using the equation:

        \begin{equation}
        F_{flash} = F_{ref} \cdot 10^{\frac{m_{ref}-m_{flash}}{2.5}}
        \end{equation}
        where $F_{ref}$ is the flux density of a reference star in photons/second/m$^{-2}$\textmu m$^{-1}$, $m_{ref}$ is the magnitude of said star, and $m_{flash}$ is the magnitude of the flash for that frame. This can then be used to calculate the power for each frame:

        \begin{equation}
            P = F_{flash} \cdot \pi f \Delta\lambda D^2
        \end{equation}
         where F$_{flash}$ is the flux density of the flash in Wm$^{-2}$\textmu m$^{-1}$, $f$ is a unitless parameter denoting the isotropy of the flash, taken here as the typically used $f$=2 as the light originated from the lunar surface~\citep{suggs2014}. $\Delta\lambda$ is the bandwidth of the observation in \textmu{m}, and $D$ is the Earth-Moon distance at the time of the flash in m. By integrating for the duration of the flash, E$_{lum}$ is calculated to be E$_{lum}$=2.12~MJ.
        
        As LIFs are modelled as black bodies, this allows the temperature to be estimated using the colour index of the flash. Because the visible observation is unfiltered light, this makes obtaining an accurate colour index difficult. Since the peak of the quantum efficiency of the camera is in the V-band, we elected to assume that Mag$_{vis}\approx$Mag$_V$. This allowed the V-J colour to be obtained simply by subtracting the J-band magnitude from the V-band magnitude of the flash in the co-occurring frame. The effective temperature is then obtained from a V-J colour index, giving the flash a peak temperature of approximately 4200~K. This value, although very uncertain due to non-synchronisation of the V and J-band observations, is reasonably inside the temperature distribution that was produced by dual band NELIOTA observations \citep{avdellidou2021}.

\section{Discussion}\label{sec:discuss}

    Here we have presented a prototype instrument for observing LIFs in the SWIR. The clear advantage of SWIR compared to visible is the increased signal due to typical LIFs temperatures, increased SNR due also to the darker sky in the SWIR compared to visible observations. The latter allows us to perform daylight observations strongly increasing the amount of time for LIF monitoring compared to classical systems working in the V-, R-, or I-bands.

    We would like to note that our prototype instrument has been tested at low altitude above sea level in an urban environment which could have contributed to decreased performances observed with respect to an ideal telescope installed in higher altitude sites. The Skywatcher flex tube is designed as a collapsible dobsonian, which extends and retracts on three metal arms. As there are no cross struts between these arms, and the weight of the camera and auto-focus assembly combined is $\approx$2~kg, the alignment of the primary and secondary mirrors might change between zenith and horizontal. This means that no matter at what altitude the telescope is in when aligned, throughout extended observation sessions (which are typical for LIF monitoring) coma might be present in at least some of the data obtained. This coma spreads the flash signal over a greater area, therefore decreasing the SNR.  

    Due to the location of the telescope, lunar observations take place partially over the city of Nice; The city can contribute both light pollution, and air pollution, which negatively affect the background light level and the atmospheric extinction. The rest of the observations take place directly over the Mediterranean sea, which contributes moisture to the atmosphere which too worsens the atmospheric extinction.

    Despite these issues that could have negatively effected the sensitivity of the methods, the technique of monitoring LIFs in the SWIR presents a unique opportunity to observe LIFs during times not available to conventional methods. Importantly, the amount of potential hours that observations can be performed is greatly increased compared to nighttime LIF monitoring by almost fivefold.
    
    There are however some improvements that could be made to increase the performance and ability to detect LIFs. As can be seen in Tab.~\ref{tab:obs}, all the LIF candidates came from observations of 10~msec or longer, despite observing at lower exposures. It is likely that this is due to the lower exposure observations taking place during the daytime, when the background is higher from both the illuminated sky and the leaking filter. Another factor which could contribute is the 10.2~msec readout time of the camera, which consequently causes exposures less than this time to be not observing over 50\% of the time. At an exposure time of 1~msec, 91\% of the light is lost to the readout time. Both of these issues could be counteracted by decreasing the aperture of the telescope, for example by using a diaphragm in front of the aperture during daylight observing. This diaphragm could then be removed for twilight- and nigh-time observing. This would reduce the amount of light entering the camera, and therefore allow for longer exposure times while maintaining the same background counts, or conversely decrease the background counts, thereby increasing the SNR, for the same given exposure.  Another possibility is to co-add several frames during the daytime observing as discussed in section~\ref{sub:stars}. 
    
    We have shown that our instrument has limiting J magnitudes of about 4.8 during daylight, from 7.8 to 8.5 during twilight and 9.3 during the night. Considering a R-J color of 1.8~mag, this would indicate that we can detect LIFs with R mag of 6.6 during the day and 11.1 during the night. Considering a luminous efficiency of $\eta_1$~=~1.5x10$^{-3}$ and $\eta_2$~=~5.0x10$^{-4}$. These correspond to energies between 1.45~GJ and 7.68~MJ for a flash with a 66~msec duration. 
    
    While the fainter events may not be detected during the day, for performing further science with the observed LIFs more energetic impacts are more scientifically rich. LIFs with a R-band magnitude >+8.0 are typically to low enough energy that the formed impact crater is near or below the resolvable limit for the LRO NAC, which has a pixel scale of approximately 0.5~m~pix$^{-1}$. For the purposes of lunar crater location, higher energy and therefore larger ejecta blankets are highly desirable for aiding in the detection of the craters, as well as for minimising the error in measurements due to being close to the resolution limit. 

    The increased duration of the J-band flash in comparison to the observed visual flash is also advantageous for LIF observations, as it presents a greater window in which the camera exposure and flash can co-occur. This thereby increases the probability for an event to be detected, as well as allowing more frames to be captured for a given flash. These both provide a more accurate measurement of the evolution of the flash, allowing for more accurate energy calculations, and for shorter events gives the possibility of more than a single frame to be captured, allowing for self confirmation. By fitting the function 
    \begin{equation}
     y = a\cdot exp\left(\frac{-x}{\tau}\right)+c
    \end{equation}
    \noindent to the observed flash in both J-band and visible, we can obtain the time constant, \texttau, for the flash decay. We found that for J-band, $\tau_J$=0.189, and for visible, $\tau_{vis}$=0.0749.

\section{Conclusions}
In this work we have presented the theoretical basis for SWIR LIF observing, and the increase in observation opportunity it allows. Moreover, we have documented the first detection of a LIF in the SWIR that was confirmed by simultaneous visible observations.

\section*{acknowledgements}
This work was supported by the Programme National de Planetologie (PNP) of CNRS/INSU, co-funded by CNES and by the program “Flash!” and co-funded by IPGP Internal research projects program. DS, MD and CA also acknowledge support from ANR “ORIGINS” (ANR-18-CE31-0014) and PL acknowledges Idex Paris Cit\'e (ANR-18-IDEX-0001).

We would also thank Luke Nicholls and QD-UKI for the loaning of demonstration equipment in 2021, which enabled the early stages of this work.

\section*{Data Availability}

The impact flash data used within this paper is available upon email request from DS (djs22@aber.ac.uk).



\bibliographystyle{mnras}
\bibliography{references} 




\appendix
\section{Description of telescope encoders}

The absolute encoders used for this project are based on the simple concept of comparing an image of a mobile scale from a fixed camera. The latter is implemented with a low-cost commercial USB microscope, which illuminates the scale by means of its LED headlights. The scale is made by vertical black-and-white lines printed in a photographic adhesive paper. The positions of the vertical bars are randomly extracted from a uniform distribution between zero and the scale length. Hence, any location on the scale is different than any other one. The camera sensor’s column are approximated aligned with the lines of the scale. The columns of each image taken by the USB camera are summed resulting in a 1D array of pixels, which is cross correlated with a template of the scale. The position of the maximum of the cross-correlation indicates the position (sub-pixel accuracy of about 1/100 can be reached when some 100 lines are co-added and spacial smoothing filtering is applied) of the centre of the image along the scale. As the scale moves, said position changes. Slight misaligned of the scale’s lines compared to the camera column is not a problem, resulting in broadening of the 1D line profile. Sub-pixel accuracy can be reached, by fitting a parabola on the magnitude of the cross-correlation function around its maximum and determining the vertex of the parabola.  When the scale is wrapped around and glued onto a disk, angular displacements of the latter, corresponding to linear displacement of the scale with respect to its image template, can be measured. It is hence possible to measure the angle of the disk.

The scale template is constructed from the same images taken with the USB microscope as follows: this procedure begins by defining a very long, e.g., 50,000 pixels, 1D empty template and a-same-length scoring template of integers. The scale and scoring template are initially equal to zero for all pixels. Next, a scanning of the entire scale by the USB microscope camera is performed. Several images per second are taken, while the scale is displaced by a small amount between each image in the positive direction. The sum of the column of the first image is added to the scale template by shifting its centre to image-width divided by two and one is added the corresponding pixels of the scoring template. The sum of the columns of each next image are cross correlated with the sum of the columns of the previous images and an integer shift is between image n and image n+1 is calculated. The sum of column of the image n are thus added to scale template. When the template of the entire (or usable section thereof) has been acquired, it is divided by the scoring 1D array (in order to calculate a mean) and the result is saved in a file.

The operation of the encoder typically consists in taking images of the scale with a typical frequency between 10 and 25 Hz from the USB microscope (using the openCV python3 framework). After some processing (spacial low-pass filtering), the sum of column of the image is cross correlated with the scale template and the position of the maximum is used to determine the centre of the image along the scale. This procedure returns a scalar position in pixels (and fraction thereof) along the scale. In order to transform pixel to sky coordinates a calibration of the scale is then performed. 

The scale calibration on the sky consists in determining an appropriate transformation that allows one to convert encoder pixel in coordinates on the sky and vice versa. This is achieved by taking images of the sky with the telescope and using a plate solving algorithm (solve-plate from astrometry.net) to determine the sky coordinates of the centre of the field. These coordinates are corrected for the precession and transformed to local hour angle ($\tau$) and declination ($\delta$). The encoder position during the acquisition of each image are also recorded, such that for each image taken on the sky there is a correspondence between sky coordinates ($\tau$, $\delta$) and the two encoder pixel positions (x, y). Finally global transformations from (x,y) to \texttau{ }and another transformation between (x,y) and \textdelta{ }are expressed in terms of a smoothing splines. This allows us to convert encoder position into equatorial sky coordinates. The solution is stable over months of operation and allows to point the telescope globally on the sky within approximately 1 arcmin, despite the high flexure of the tube. Possible improvement is to mount the optics within a more rigid tube (e.g. carbon fiber) or truss.  


\bsp	
\label{lastpage}
\end{document}